# Semantic Integration & Single-Site Opening of Multiple Governmental Data Sources

**Konstantinos I. Kotis**
(Ai-Group, Dept. of Digital Systems, University of Piraeus, Piraeus, Greece
and
(North Aegean Regional Administration (NARA), Samos, Greece
kotis@aegean.gr)

**Iraklis I. Athanasakis**
(North Aegean Regional Administration (NARA), Samos, Greece
h.athanasakis@samos.pvaigaiou.gov.gr)

**George A. Vouros**
(Ai-Group, Dept. of Digital Systems, University of Piraeus, Piraeus, Greece
georgev@unipi.gr)

**Abstract:** In many cases, government data is still 'locked' in several 'data silos', even within the boundaries of a single (inter-)national public organization with disparate and distributed organizational units and departments spread across multiple sites. Opening data and enabling its unified querying from a single site in an efficient and effective way is a semantic application integration and open government data challenge. This paper describes how NARA is using Semantic Web technology to implement an application integration approach within the boundaries of its organization via opening and querying multiple governmental data sources from a single site. The generic approach proposed, namely S3-AI, provides support to answering unified, ontology-mediated, federated queries to data produced and exploited by disparate applications, while these are being located in different organizational sites. S3-AI preserves ownership, autonomy and independency of applications and data. The paper extensively demonstrates S3-AI, using the D2RQ and Fuseki technologies, for addressing the needs of a governmental "IT helpdesk support" case.

## 1    Introduction

It has been argued [Bizer, 09; Ding, 12] that opening and linking data on the Web (delivering data in both machine and human-readable forms) is expected to promote transparency and collaboration between governments (G2G), governments and citizens (G2C), governments and business (G2B), allowing the creation of new, smarter, innovative, added-value public services and applications, towards realizing more effective and qualitative decision-making processes. Governments, businesses and citi-zens may develop new applications and services in order to work with, analyze, and make sense of the linked open government data (LOGD).

In many cases today, government data is still 'locked' in several 'data silos', even within the boundaries of a single (inter-)national public organization with disparate and distributed organizational units and departments spread across multiple sites. The aim of the LOGD initiative [Ding, 12], aligned with the European Union PSI directive[1] for reusing public sector data, is to encourage and support public administrations to adopt open data policies and to enable efficient application integration via the opening and virtualization of multiple governmental data sources, without requiring to redesign information systems and/or to store data in centralized 'data silos'. At the same time, the initiative supports the preservation of autonomy and ownership of the original data, meaning that linking between data is established with owners to still have full control of their original data and application (which may be also preserved in its original form) [Bizer, 09].

This paper presents a system for the opening of data stored in multiple repositories across an organization and enabling its unified ontology-mediated querying from a single organizational site, in an effective and efficient way. This is seen as a semantic application integration and linked open government data (LOGD) challenge for the North Aegean Regional Administration (NARA). Application integration in this context, closely related to data integration, aims at integrating 'live' and evolving data collections created by applications in an un-coordinated way, at real-time.

NARA is a European regional islander public administration authority located in the Aegean Sea and spread in five different sites (islands). The paper describes how NARA is using Semantic Web technology to support application integration via an efficient and effective approach of opening and querying multiple governmental data sources from a single site. The implementation of the developed approach is currently limited to the opening of data within the boundaries of the organization and allows disparate applications and their data located in different organizational sites to preserve their autonomy and independency, and at the same time to provide support for executing ontology-mediated federated queries from a single site.

The core technology used in the system is a) the D2RQ open source software for accessing relational databases as virtual read-only RDF graphs [Bizer, 07; Cyganiak, 06] and b) the Fuseki open source server for efficiently executing federated ARQ SPARQL 1.1 queries over multiple sources [Fuseki, 13; Montoya, 12]. For presentation purposes the paper discusses this framework under the working label 'Single-Site Semantic Application Integration' (S3-AI).

The presented work contributes to the following areas of research: a) the use of semantic technologies to ease the querying of data from different sources through a single point of access, and b) the use of common/reference vocabularies to enable the opening and linking of data from different sources towards efficient application integration. In a more 'in-use' oriented view of SW research, the presented work contributes to the following topics: a) Implementation of the generic S3-AI approach and the use of semantic technologies for opening and integrating data from disparate e-Government applications, b) Reporting on lessons and best practices from deploying and using the S3-AI approach in a specific e-Government domain, and c) investigat-

---

ing the pragmatics of using and deploying semantic technologies towards integrating multiple governmental data sources.

This paper is structured as follows: Section 2 describes the motivating points and the application setting of the S3-AI approach. Section 3 presents the overall approach while section 4 presents the technologies and the IT helpdesk support case for deploying S3-AI, as well as the vocabularies that have been used for the description of data in the specific e-Government domain. Section 5 discusses results of the evaluation task performed recently. Section 6 presents work closely related to the described approach and its objectives. Section 7 discusses lessons learned from the use of the approach in NARA e-Government setting and section 8 concludes the paper with a reference to our future work.

## 2 Application Integration Setting and Requirements

NARA is a European regional islander administration authority located in the North Aegean Sea and spread in five main geographical sites, each site corresponding to an island with an individual governmental organizational unit. Each of the five organizational units is a formal member of NARA, under a semi-distributed form of administration (administrative autonomy of units is partially preserved). NARA is a new formal organization shaped after a Greek government reform initiative in 2010, merging three former prefectural authorities in North Aegean, i.e. Samos Prefecture, Lesvos Prefecture and Chios Prefecture. Merging of organizational units, such as the merging of prefectural IT departments, resulted to a form of semi-autonomy: There are common general policies applying to them but they do follow local, specialized and focused policies (mainly due to employees' expertise and local administration needs). Such an organizational reform, at the administration level, resulted to a similar situation at the technical level, where several contracted and outsourced IT applications used for G2C and G2G services have been redesigned and adapted to new integral needs e.g. BackOffice applications. Nevertheless, open source and free/libre software re-used by IT departments' management purposes, have not been considered for integration until recently (due to policy and decision-making reasons).

This paper describes the design and development of the S3-AI approach, towards the effective and efficient integration of data sources maintained by different applications from a single organizational site in a unified manner. The main requirements that an integration approach must fulfil towards this target are:

- It must be easily deployed within NARA and other governmental organizations, in different domain-specific environments, and by IT employees with different levels of expertise.
- It must be easily deployed with the minimum cost and effort, while at the same time being customizable to specific needs and requirements concerning the management of data. Thus, it is preferable to use open-source technological components, while allowing the use and integration of other technologies, towards fulfilling specific needs and requirements.
- It must not impose significant maintenance effort additional to the one already necessary for the maintenance of the integrated applications/datasets.

- It must preserve ownership, autonomy and independence of original/source applications and data.
- It must support the automatic and continuous integration of 'live' and evolving operational data created by multiple applications in real-time.

The S3-AI approach has been deployed and evaluated in the real-world case for the integration of helpdesk support applications used in different NARA sites. This case concerns the IT support ticketing dataset recorded for Samos regional unit IT department using osTicket open source ticketing software[2], and the ticketing dataset recorded in the Ikaria regional unit using GLPI open source IT and Asset Management software[3]. The availability of open data is currently limited within the boundaries of the organization and among the different NARA-based IT departments' employees. For presentation purposes open data are filtered due to privacy and security issues (descriptions of tickets may contain employees' names, e-mail addresses and login information).

## 3 The Overall S3-AI approach

NARA followed the Linked Open Government Data paradigm [Ding, 12] towards meeting the design requirements of an application integration approach. Relational database sources are transformed to virtual RDF graphs by utilizing RDB to RDF mappings (one-to-one correspondences between RDB tables/columns and RDF classes/properties), based on specific rules of D2RQ Mapping Language. As depicted in the general architecture of the designed approach in Figure 1, an RDB2RDF transformation process initially generates RDB2RDF mapping files for each RDB that records data of the applications to-be-integrated (1st level mapping).

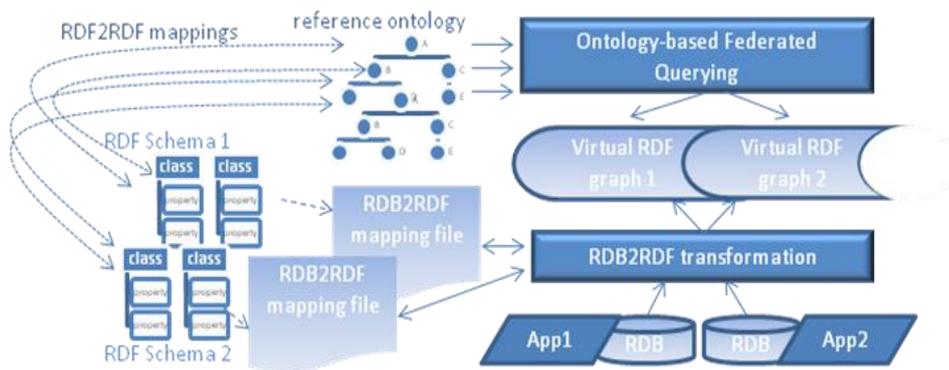

*Fig. 1. Architectural design of S3-AI approach*

This is automatically done using a default, auto-generated RDF schema (following the rule of one-to-one correspondence from RDB to RDF). This RDF schema is then mapped to a reference ontology (2nd level mapping) in order to identify similarities between schema elements and the elements of the reference ontology. The aim is to use the corresponding elements of the reference ontology in the RDB2RDF mapping files so that data are described (as much as possible) using a shared and agreed vocabulary with well-defined semantics, following the guidelines of the LOD paradigm. Using the updated RDB2RDF mapping files, virtual RDF graphs are then created: No persistence storage is required. These graphs are then accessible in the Web via corresponding SPARQL endpoints.

Several RDBs can be RDFized by running distinct RDB2RDF processes in parallel. Ontology-mediated queries (using the elements of the reference ontology) are then routed (via a federation process) to the specific SPARQL endpoints and their results are unified/combined at the interface level.

Beyond the obvious advantage of opening data and sharing knowledge within and across governmental organizational bodies, the advantages of the S3-AI integration approach can be summarized in the following points:

- The autonomy, independence and ownership of original applications and their generated data is preserved. Data, although opened and integrated, can still be exploited independently, by means of the original applications, or by other applications and services. Data sources and the related applications are preserved in their original RDB form.
- The single-site S3-AI architecture does not require stakeholders, at the sites where applications to-be-integrated operate, to take any further action for maintenance beyond the one already put for maintaining and managing their own applications and the respective data.
- Updating the auto-generated RDF schemas with terms from shared and agreed RDF vocabularies (e.g. W3C-recommended or well-known and publicly accessible ontologies), allows Semantic Web client applications to exploit more of the opened data.
- The approach, based on the process of auto-generating virtual RDF graphs, supports the automated and continuous integration of 'live' and evolving operational data created by multiple applications in real-time. Most importantly, this can be done in a relatively simple way.
  Last but not least, the approach does not require the use of specific proprietary technologies but allows for highly customizable implementations using a variety of alternative SW-related open-source software (RDB to RDF mapping technologies, SPARQL servers, etc).

## 4    S3-AI for the 'eGov IT Helpdesk Support' case

In this section we present the main building technologies of the S3-AI implementation, namely the D2RQ technology and the Fuseki Server, and how these technologies were used towards implementing of the S3-AI for eGov IT helpdesk support in

NARA. The selection of the building technologies for this implementation was driven by the abovementioned S3-AI requirements.

### 4.1 D2RQ and Fuseki technologies

D2RQ is a simple to install and use free/libre (Apache Licence) open-source software platform for accessing relational databases as virtual (read-only) RDF graphs. It provides RDF-based access to the content of a relational database without having to replicate it into an RDF store, thus avoiding additional implementation and maintenance efforts. Among other things that D2RQ supports, it is possible to: a) remotely access a D2RQ-mapped database via the SPARQL protocol (SPARQL endpoint) b) access RDF descriptions of individual entities in the database by dereferencing their URIs (Linked Data provides RDF links as an alternative to registered-based discovery mechanism for the Semantic Web), and c) view content in HTML (for assisting in writing and debugging mappings). One thing that D2RQ cannot do (since mapping relational databases to RDF is a local problem) is integration of multiple databases or other data sources. For further information please access its web documentation at http://d2rq.org/.

Expressive queries over multiple data sources can be answered either by replicating data in a local repository (RDF store), or by using a query federation service to sent query fragments to multiple data sources (and then integrate answers to obtain the required result). Both directions seem to be currently active in the research community. NARA follows the latter to meet the requirements of S3-AI approach. A free/libre open-source ARQ SPARQL 1.1 query engine integrated in the Fuseki server (Apache License) has been used. A latest W3C specification[4] defines the syntax and semantics of SPARQL 1.1 federated query extension for executing queries distributed over different SPARQL endpoints. The SERVICE functionality extends SPARQL 1.1 to support queries that merge data distributed across the Web. Fuseki is a SPARQL server capable of REST-style SPARQL 1.1 query using the SPARQL protocol over HTTP. For further information please access its web documentation at http://jena.apache.org/documentation/serving_data/.

### 4.2 The eGov IT Helpdesk support system

Using the D2RQ technology, relational databases are transformed to virtual RDF graphs by utilizing auto-generated RDB to RDF mapping files (i.e. using the correspondences generated by the 1st level mapping). The RDB2RDF transformation functionality of D2RQ initially generates an RDB2RDF mapping file in Turtle notation by simply taking as input the URL of the relational database and accessibility details (username, password and JDBC driver). The generated mapping file is then used by the D2R server to create on-the-fly the virtual RDF graph and make it accessible via a SPARQL endpoint, at a specific URL and port. Different RDBs can be RDFized at different ports by running as many D2R server instances.

---

NARA follows a single-site approach for that: As depicted in Figure 2, we present the integration of two distinct applications in the case of eGov IT helpdesk support: S3-AI uses two D2R server instances (sites 1 and 2) on a single integration site (site 3: the S3-AI server). This way, all the opened connections of the S3-AI implementation to multiple RDBs are managed from this single site, preserving RDBs at their original location and generating independent and separate RDF graphs (one graph for one RDB) at the integration site.

As already stated, to open data and access them in a unified manner from a single site, shared and commonly agreed vocabularies are required for describing them. The S3-AI eGov IT helpdesk support system uses a domain-specific reference vocabulary for semantically annotating the data to be integrated. If such an ontology does not exist, the involved stakeholders need to develop one, reusing also terms from other public and widely-used related vocabularies (found in public vocabulary services such as the Linked Open Vocabularies - LOV).

The computation of mappings (2nd level mapping) between terms of the reference ontology and the individual D2RQ auto-generated RDB2RDF mapping files might be a time-consuming and error-prone task if not carefully executed. (Semi-)automated schema/ontology mapping tools, depending on their performance and on the nature of the source ontologies, can influence either positively or negatively the performance of an S3-AI implementation. On the other hand, devising a lightweight reference vocabulary from scratch to reflect a) end-users' requirements for querying their data in a unified manner, and b) obvious commonalities of RDF schemas, can improve the performance of the mapping process and consequently the overall performance of the S3-AI system. The latter is the approach that we have followed in deploying the S3-AI system for the NARA 'eGov IT helpdesk support' case.

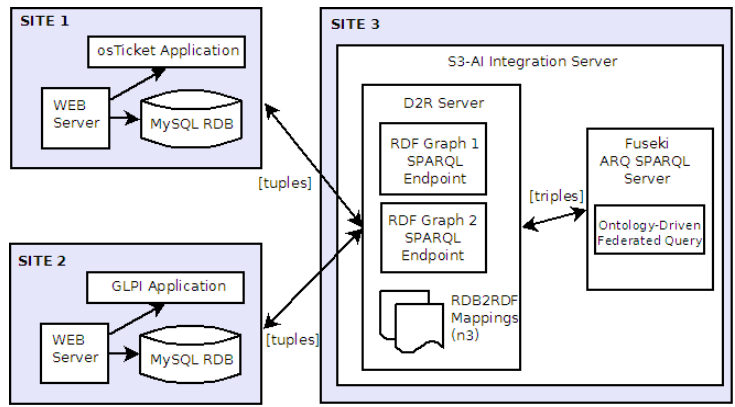

*Fig. 2. Topology of an S3-AI implementation for integrating two sites*

The S3-AI can easily extend its range towards integrating multiple applications by simply a) creating new D2R server instantiations, and b) exploiting the mappings between the auto-generated RDF vocabularies and the reference ontology used, and use the reference ontology terms within the RDB2RDF mapping files.

As already pointed out, the main building blocks of the S3-AI system can be replaced by technological components that meet further requirements: E.g. components that scale up better than those used in the NARA's eGov IT helpdesk support implementation. For the presented case the selected technologies i.e. D2RQ and Fuseki server, were more than adequate (as it was also verified by the evaluation results of this work).

Finally, the implemented system supports the automated and continuous integration of 'live' and evolving IT helpdesk support data created by the multiple applications in real-time. With D2RQ technology, SPARQL requests are rewritten into SQL queries that are sent to the original data sources via the pre-computed RDB2RDF mappings. This on-the-fly translation allows publishing of RDF from large live databases and eliminates the need for replicating the data into a dedicated RDF triple store.

### 4.3 Semantic Annotation of Data

The S3-AI approach uses a set of vocabularies to semantically describe and further retrieve data. In the following paragraphs the paper briefly describes those and states their role in the S3-AI eGov IT helpdesk support system. NARA follows the W3C definition of a vocabulary and the guideline of how it can be used [W3C, 13a].

The D2RQ mapping language is a declarative language for describing the relation between a relational database schema and an RDF schema (or an OWL ontology). A D2RQ mapping is itself an RDF document written in Turtle syntax[5]. The mapping is expressed using terms in the D2RQ namespace (http://www.wiwiss.fu-berlin.de/suhl/bizer/D2RQ/0.1#). The terms in this namespace are formally defined in the D2RQ RDF schema[6]. The 'generate-mapping' tool of the D2RQ platform automatically generates the D2RQ mapping file by analyzing the schema of an existing database.

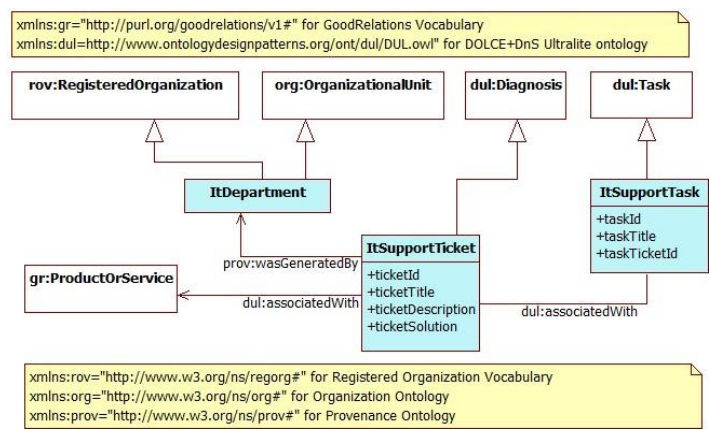

*Fig. 3. The helpDeskOnto reference ontology abstract design*

The 'Core Business Vocabulary', developed by ISA [EC-ISA, 13] and recently renamed by W3C GLD WG to 'Registered Organization Vocabulary' (RegOrg) [W3C-ROV, 13] is a vocabulary for describing organizations that have a legal entity status (formally registered organizations in a national or regional register). It captures the fundamental characteristics of a legal entity, e.g. its legal name, its registered activities and address, and it can be used to describe private and governmental organizations. The 'Organization' vocabulary (ORG), also recommended by W3C GLD WG [W3C-OrgOnto, 13] is designed to enable publication of information on organizations and organizational structures including governmental organizations. It is intended to provide a generic, reusable core ontology that can be extended or specialized for use in particular situations. The ORG vocabulary has been already used by the UK Government for describing public organizations. The ORG vocabulary is more generic than the RegOrg. Both vocabularies abide by the Linked Data principles [EC-ISA-JoinUp, 13a]. We have reused both vocabularies to classify the classes and properties of the 'HelpDeskOnto' domain-specific reference ontology, based on the argument that if Semantic Web applications can interpret a generic property (super-property) or class (super-class) are then able to interpret the data described by the more specific properties or classes [EC-ISA-JoinUp, 13b; W3C, 13b].

Following LOD best practices and specific guidelines from W3C, ISA and LOV cookbooks [EC-ISA-JoinUp, 13b; W3C, 13b; Vandenbussche, 13] we have devised a domain-specific reference ontology to represent knowledge about the 'eGOV IT helpdesk support' domain. The developed namespace of the reference ontology (figure 3) is *xmlns:hdo = http://www.samos.gr/ontologies/helpdeskOnto.owl#*.The main elements of the ontology are described below:

— The class *hdo:ItDepartment:* is defined as a '*rov:RegisteredOrganization*' and a '*org:OrganizationalUnit*'
— The class *hdo:ItSupportTicket*: is defined as a '*dul:Diagnosis*' i.e. 'a description of the situation of a system, usually applied in order to control a normal behaviour, or to explain a notable behaviour e.g. a functional breakdown'
— The class *hdo:ItSupportTask*: : is defined as a '*dul:Task*'
— The class '*hdo:ItSupportTicket*' is '*dul:associatedWith*' class '*hdo:ItSupportTask*' and '*gr:ProductOrService*'.
— The class '*hdo:ItSupportTicket*' '*prov:wasGeneratedBy*' class '*hdo:ItDepartment*'

As depicted in Figure 3, the *hdo* namespace defines only a few classes and properties that are currently adequate for an S3-AI first implementation in the eGov IT helpdesk support case. This application domain, to the best of our knowledge, has not been researched before in the context of semantic application integration and LOGD, and the domain knowledge has not been explicitly represented before using a formal vocabulary. The developed OWL-encoded reference ontology is validated and published (for reuse, agreement and consensus purposes) by public vocabulary services such as LOV [LOV, 13] and Joinup/ISA [EC, 13] (researchers of both services have been contributed towards a well-defined namespace, following their specific guidelines) at http://lov.okfn.org/dataset/lov/details/vocabulary_hdo.html and https://joinup.ec.europa.eu/catalogue/asset_release/it-helpdesk-support-ontology respectively.

### 4.4 S3-AI system deployment

This section outlines the tasks performed for the deployment of the S3-AI approach in the case the eGov IT helpdesk support system application integration. As depicted in the case of Figure 2, NARA integrated two heterogeneous applications running on two different sites of the organization, using different databases (heterogeneous schemas) to store the related data, with the aim to execute common queries such as "*Get me tickets (from any site) that are about "No Video" problem*" or "*Get me solutions (from any site) for tickets that were about "No Video" problem*". The RDB schemata of the data sources for the IT helpdesk support ticketing, are quite different mainly due to the additional 'asset management' functionality of the GLPI application. Both applications store data at local MySQL databases and are accessible via a Web interface.

To develop the eGov IT helpdesk support system using the S3-AI approach, one must perform the following tasks:

— Task 1: Download and install the D2R server software (version 0.8.1) on a Linux machine (Site 3 in Figure 2).
— Task 2: Since drivers for MySQL are already included in this distribution, create the two auto-generated mapping files, one for each source (under-integration) site (Site 1 and Site 2 in Figure 2) by just executing the 'generate-mapping' tool of the D2RQ distribution. The related Linux commands and part of the mappingOS-ticket.n3 mapping file (Figure 4) is shown below.

```
1   prefix map: <#> .
2   @prefix db: <> .
3   @prefix vocab: <vocab/> .
4   @prefix rdf: <http://www.w3.org/1999/02/22-rdf-syntax-ns#> .
5   @prefix rdfs: <http://www.w3.org/2000/01/rdf-schema#> .
6   @prefix xsd: <http://www.w3.org/2001/XMLSchema#> .
7   @prefix d2rq: <http://www.wiwiss.fu-berlin.de/suhl/bizer/D2RQ/0.1#> .
8   @prefix jdbc: <http://d2rq.org/terms/jdbc/> .
9
10  map:database a d2rq:Database;
11      d2rq:jdbcDriver "com.mysql.jdbc.Driver";
12      d2rq:jdbcDSN "jdbc:mysql://10.129.46.217/osticket10";
13      d2rq:username "XXXXXXXX";
14      d2rq:password "XXXXX";
15      jdbc:autoReconnect "true";
16      jdbc:zeroDateTimeBehavior "convertToNull";
17      .
18
19  # Table ost_ticket
20  map:ost_ticket a d2rq:ClassMap;
21      d2rq:dataStorage map:database;
22      d2rq:uriPattern "ost_ticket/@@ost_ticket.ticket_id@@";
23      d2rq:class vocab:ost_ticket;
24      d2rq:classDefinitionLabel "ost_ticket";
25      .
```

*Fig. 4. Part of the auto-generated mapping file using the default RDF schema, depicting the mapping of table "ost_ticket" to a class "ost_ticket" (lines 20-24) in the "vocab" namespace*

- #for osTicket

  ```
  ./generate-mapping  -o  mappingOSticket.n3  -u  xxxxx  -p  xxxxx
  jdbc:mysql://10.129.46.217/osticket
  ```

- #for GLPI

  ```
  ./generate-mapping  -o  mappingGLPI.n3  -u  xxxxx  -p  xxxxx
  jdbc:mysql://10.129.46.218/glpi
  ```

— Task 3: Start two instances of the D2R server on the main (integration) site (i.e. site 3 in Figure 2), one for each mapping file, using different network ports. Use the default 2020 port for the first instantiation and assign port 2021 for the second instantiation of the D2R server. The related Linux commands are:

- ./d2r-server -b http://eupalinos.samos.gr:2020/ -p 2020 --fast mappingOSticket.n3
- ./d2r-server -b http://eupalinos.samos.gr:2021/ -p 2021 --fast mappingGLPI.n3

— Task 4: Test the first S3-AI steps by individually browsing data and using the SPARQL explorer to execute queries and display results in different formats per site (Figure 5).

*Fig. 5. RDF browsing (left) of "No Video" ticket with resource ID 1149 in the ost-Ticket dataset using Snorql SPARQL explorer. Other views (right) are also available using either HTML or Ajax-based SPARQL explorer.*

— Task 5: If there is not any agreed reference ontology, then develop a new one, driven by the requirements of the application integration i.e. by a set of representative queries that users need to run against the integrated data. For this deployment, we have developed a simple ontology (Figure 3), discussed in section 4.3.

— Task 6: Customize the two auto-generated RDB2RDF mapping files using the reference ontology.

```
1  prefix map: <#> .
2  @prefix db: <> .
3  @prefix vocab: <vocab/> .
4  @prefix rdf: <http://www.w3.org/1999/02/22-rdf-syntax-ns#> .
5  @prefix rdfs: <http://www.w3.org/2000/01/rdf-schema#> .
6  @prefix xsd: <http://www.w3.org/2001/XMLSchema#> .
7  @prefix d2rq: <http://www.wiwiss.fu-berlin.de/suhl/bizer/D2RQ/0.1#> .
8  @prefix jdbc: <http://d2rq.org/terms/jdbc/> .
9  # Namespace of the helpdeskOnto ontology
10 @prefix hdo: <http://www.samos.gr/ontologies/helpdeskOnto.owl#> .
11
12 map:database a d2rq:Database;
13     d2rq:jdbcDriver "com.mysql.jdbc.Driver";
14     d2rq:jdbcDSN "jdbc:mysql://10.129.46.217/osticket10";
15     d2rq:username "XXXXXXXX";
16     d2rq:password "XXXXX";
17     jdbc:autoReconnect "true";
18     jdbc:zeroDateTimeBehavior "convertToNull";
19         .
20
21 # Table ost_ticket
22 map:ost_ticket a d2rq:ClassMap;
23     d2rq:dataStorage map:database;
24     d2rq:uriPattern "ost_ticket/@@ost_ticket.ticket_id@@";
25     d2rq:class hdo:ItSupportTicket;
26         .
```

*Fig. 5. Part of the updated mapping file using the helpdesk (hdo) ontology schema, depicting the mapping of table ost_ticket to the new class "ItSupportTicket" (line 25) in the "hdo" namespace*

— Task 7: Download and install the Jena-Fuseki SPARQL Server (version 0.2.6) on a Linux machine (default port is 3030) in order to run the federated queries, addressing them to both SPARQL endpoints (Figure 6).

In Task 6, we followed a simple approach by replacing the terms in each of the two mapping files generated by D2RQ, with the corresponding terms used for lexicalizing classes and properties in the *helpDeskOnto* reference ontology developed. We decided not to use a schema/ontology mapping tool to assist the mapping task, although we had access to such tools, due to the low complexity of the mapping case. We have also developed a simple Web interface for accessing the opened data sources, via HTML and SPARQL, and for executing ontology-mediated federated queries over integrated sources. For reviewing and demonstration purposes we have made publicly available two individual datasets and a federated query endpoint, as well as example queries from the http://www.samos.gr/apps/s3-ai/eGovTicketApp.xhtml Web page (directly accessing them from http://eupalinos.samos.gr:2020/, http://eupalinos.samos.gr:2021/ and http://eupalinos.samos.gr:3030/sparql.tpl respec-tively). The work is ongoing and thus we expect in the near future to integrate additional datasets of the same domain, from other NARA-based organizational units or external ones.

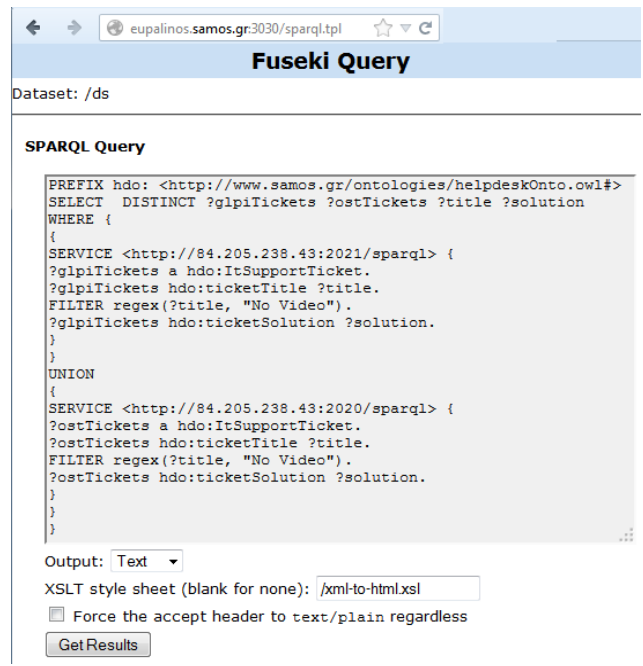

*Fig. 6. The "No Video" example SPARQL query executed at http://eupalinos.samos.gr:3030/sparql.tpl*

## 5    Evaluation of the S3-AI for eGov IT helpdesk support

For the evaluation of the S3-AI system used in the context of the eGov IT helpdesk support integration system, beyond internal interviews with IT engineers, we provided a simple Web interface for IT employees. Multiple ways of accessing and querying the data were provided:

- Browsing and querying data of each individual site, by providing a SPARQL endpoint, an RDF view and an HTML view for each dataset
- Executing predefined SPARQL 1.1 federated queries for the retrieval of data,
- Executing ontology-mediated SPARQL 1.1 federated queries, using the semantics of the reference ontology to form the fields querying templates.

The users in this use case are 8 IT employees at different levels of expertise, i.e. managers, engineers or technicians, in three IT departments, having their own experience on different IT helpdesk support applications, using the system since May 2013. During an internal interview, users made the following qualitative comments:

- The S3-AI approach, from IT engineers and managers view, is meeting the challenge for a customizable and simple (but not simplistic) application integration approach.
- The autonomy of the original applications and the preservation of the data sources are of a major importance to IT managers, and S3-AI approach delivers this to a great extent.
- The capability of accessing and querying each other's datasets in multiple views (HTML, RDF, SPARQL endpoints), individually or in a unified manner, is a valuable asset for analyzing, understanding, and sharing data and knowledge across the organization, overcoming the boundaries of local 'data silos' .
- The use of free, open-source technologies for the implementation of the S3-AI approach, resulting to a low-spending e-government project, is of a great importance also for managers and decision-makers. The low-cost character of the approach allows for a realistic planning of future e-government application integration projects, based on the S3-AI approach.
- In terms of the ontology development and mapping tasks, they reported a high degree of complexity although they had been involved in the past in an ontology development task for the needs of another e-government project, namely Samos-Dialogos.gr [Anadiotis, 10]. They reported the need to be assisted by a knowledge engineering during the ontology mapping task, but as also reported in other related work [Alani, 07] data-driven lightweight ontologies can be devised relatively easily.

It is clear that such a qualitative assessment of the s3-ai approach must be performed in other domains that might have disparate requirements for the use of data, and also by additional engineers. A range of application scenarios is clearly required and this is our future aim as far as the deployment of different S3-AI implementations is concerned.

Regarding quantitative results, we have measured the performance of the S3-AI implemented system in accordance to the number of applications/datasets that the approach integrates for the particular domain. For the purposes of the evaluation, test sites were consisting of 84.192 triples each, and we have replicated the two sites involved in our case so as to test the scalability of our approach.

More specifically, the evaluation showed (see figure 7) that:

- The memory capacity of the querying site is increasing linearly as the number of sites increases. The main reason for this is the hosting of a separate running D2R server instance for every connected site. This is a drawback of the particular implementation and could probably be resolved by either customizing the D2R software or using an alternative multi-process RDB2RDF technology.
- The computational power demanded by the querying site increases linearly, as the number of sites to be integrated, increases.

We setup an experimental server using virtualization technology in order to overcome network factors that usually influence evaluation results. More specifically, we have used the PROXMOX Virtual Environment[7] in order to create three different virtual Linux machines, each one running an individual OS: one machine for each of the two ticketing applications and one for the S3-AI application integration software. The virtual machines' (OpenVZ Containers) resources were all set at: RAM = 256MB, HDD = 4GB, OS = Ubuntu 12.04 Server 32bit.

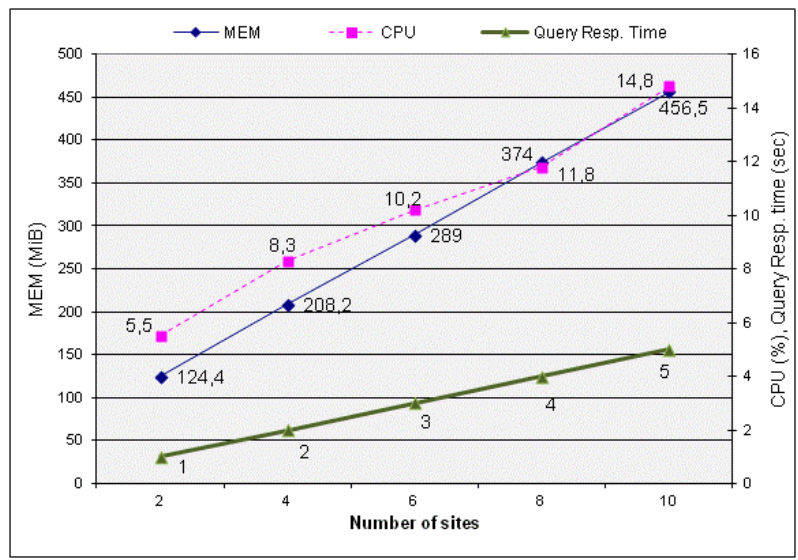

*Fig. 7. S3-AI performance measured with a number of integrated sites*

Regarding portability of the implemented system, this depends on the following issues:

- The existence of a reference ontology: If there isn't any to be reused, a new one must be developed from scratch.
- The expressiveness of the developed or reused reference ontology: While an expressive ontology is a clear advantage if its expressiveness is exploited by the system components, a reference ontology must be suitable in a way that the process of mapping its elements to auto-generated RDF schema elements is facilitated.
- The capability of IT engineers to develop simple data-driven lightweight ontologies.

---

## 6 Related Work

The interoperability solution for European public administrations (ISA) initiative [EC-ISA, 13] is carrying out several LOGD pilots in EU member states. The first pilot referenced there is about interconnecting Belgian national and regional address registers. It uses ISA's Core Location vocabulary to open and interlink data from the address registers of three Belgian regions, namely UrBIS in the Brussels Region, CRAB in the Flanders and PICC in Wallonia, and the Belgian National Geographic Institute (NGI). A demo of the pilot is available at http://location.testproject.eu/BEL/. The second pilot is called AMDS.SW and concerns descriptions from national and regional forges. This pilot aims at publishing sample descriptions of reusable, open-source software from national and regional forges in Europe in HTML+RDFa, RDF-XML, and Turtle formats using the Asset Description Metadata Schema for Software (AMDS.SW) vocabulary. So far, the pilot has contributed a number of sample descriptions of reusable, open-source software on CENATIC.es and Apache.org in HTML+RDFa, RDF-XML, and Turtle formats. The RDF data was created using the Google Refine RDF spreadsheet template. For more detail, please visit https://joinup.ec.europa.eu/asset/adms_foss/description. Both efforts are demonstrating the opening and linkage of governmental data; however they lack of an approach for their federated querying (using an ontology-mediated approach).

Closely related to opening eGov data is the approach proposed by the School of Science and Technology at International Hellenic University (IHU) in [Varitimou, 2013]. The main purpose of this work is to make information about business entities in Greece publicly available (open), by providing the descriptions of the businesses as open linked data. Data was reused from Publicspending.gr SPARQL endpoint created in the context of another Greek project (by National Technical University of Athens) that visualizes national public spending data. Core Business and Core Location Vocabularies (recently updated using RegOrg ontology) from the European Commission ISA Programme have been also used to describe the data, and an OpenLink Virtuoso platform instantiation has been used as the RDFization and triplestore technology of the system. However, these efforts have not been designed to address multiple data sources integration and federated querying.

In the past, NARA has been involved in another semantic application integration task in the eParticipation domain, in the context of a project called Samos-Dialogos.gr [Anadiotis, 10]. Although the main goal of the project was to facilitate dialogue using semantic technologies for e-participation at Samos Prefecture (nowadays merged with NARA), the data created during the evaluation of the project was opened using the D2RQ technology, mapped against a SIOC-based deliberation ontology. There are SPARQL endpoint that were made available for executing federated queries as well as for accessing data generated from other Dialogos sites in Greece. This work emphasized on the sharing aspect of the implemented platform, showed how it aligns with the Linked Data philosophy and infrastructure, and described a distributed approach to contextual views retrieval that was based on it. The dataset of NARA at Samos can be accessed from http://www.samos-dialogos.gr:2020/ endpoint.

Beyond the e-Government domain, there are other domains to report related work on semantic data integration using a LOD-based approach. Researchers of 'SWISS Experiment' work on Sensor Data and Semantic Sensor Networks domain and specifically on semantic sensor data search in a large-scale federated sensor network [Calbimonte, 11]. The relevance to this paper resides at the definition of virtual RDF streams (since data is recorded in data streams coming from multiple sensor data sources), whose ontological terms are related to the underlying sensor data schemas through declarative mappings expressed in the R2RML mapping language (for mapping relational schema to RDF), and queried in terms of a high level sensor network ontology, namely the SSN ontology.

In LOD2 project, the aim is to create knowledge out of interlinked open data and in most cases out of governmental data. LOD2 actually provides a set of tools developed as building blocks of a unified LOD framework. Differently from S3-AI, the LOD2 software stack integrates Virtuoso and SparQLed technology instead of the D2RQ and Fuseki technologies. For more on LOD2 project please visit http://lod2.eu/.

Closely related to S3-AI, SemWIQ [Langegger, 08] is a mediator-based system for virtual data integration based also on the D2R and SPARQL (Jena/ARQ2) technology. The system has been primarily developed for sharing scientific data. According to SemWIQ query federation is realized via SPARQL and the Jena/ARQ technology. Beyond our work, they discuss distributed query plans and their optimization and promise future inclusion of OWL-DL reasoning for more effective query execution. S3-AI places much emphasis on the mapping levels towards unified querying of the data sources.

Last but not least, the DataLift project [Scharffe, 12] is using a similar approach to expose and interlink data in RDF (in two steps only). The RDF extensions for Google Refine allow publication of legacy data as RDF [Mali, 12]. Both projects have no support for ontology-mediated querying of distributed SPARQL endpoints. In [Lopez, 12] a commercial and custom (using IBM technology) approach is applied in the same domain and in [Alani, 07] a similar process for relational databases is facilitated but with no support of federated querying. The findings of both works are partially aligned with the ones presented in our discussion section. Data heterogeneity and information integration problems in eGovernment environments have been also tackled from the service-oriented and agent-based side, using SW technologies, as demonstrated in the MAS system [García-Sánchez, 11].

## 7 Discussion

The S3-AI approach itself does not limit the number of applications/data sources to be integrated. This depends on the scalability of the Semantic Web technologies used for a particular deployment. One can customize the S3-AI approach by simply replacing any technological component with a more scalable one, given that the new technology provides at least the functionality demonstrated in the presented S3-AI implementation. The recommendation is to use the technologies used in the presented case (mainly due to their simplicity, free open-source character and customization capabil-ity) and replace with others, if scalability problems (or other limitations) occur during deployment.

The S3-AI approach can be used in any domain, given that a domain-specific reference ontology exists that much the data requirements. One may of course re-use an existing ontology or develop a new one. One must carefully decide whether an existing reference ontology is suitable for integrating the applications at-hand. Such a decision is not always easy to make, and it heavily depends on the schemata of the data sources to be integrated: It may lead to hard and problematic cases concerning the computation of mappings between this ontology and the auto-generated RDF schemas. Having said that, the recommendation is to reuse existing vocabularies where possible, refining them as needed, rather than developing new ones, which in most cases is not a difficult task, as reported in [Anadiotis, 10] and by UK governmental bodies [Alani, 07].

An important decision to be made during the S3-AI approach concerns the use of an ontology/schema mapping tool to assist the $2^{nd}$ level mapping in the semantic annotation process. This decision is important since it is not always effective to rely on a (semi-)automated ontology mapping tool due to a variety of reasons such as the expressiveness of the reference ontology, the accuracy of the mapping tool, the size of the vocabulary to map, to mention a few. Although there are many challenges towards realizing ontology mapping tools that meet the requirements of integration applications, a hybrid approach that uses a semi-automated mapping tool to assist the users by only suggesting the mappings between large RDF schemas and reference ontologies, is recommended. In any case however, the process of mapping terms and replacing them in the mapping files can cause problems due to a) identification of incorrect pairs of terms to be mapped, b) incorrect handle of correctly identified pairs of terms i.e. misplacement of terms in the mapping file, c) syntax errors of the placed terms in the mapping file, d) no placement in the mapping file of correctly identified pairs of terms.

## 8    Conclusions and Future Work

This paper presented the S3-AI approach and a first implementation for semantic application integration via a single-site opening of multiple governmental 'data silos'. The presented approach has been implemented using the D2RQ and Fuseki free open-source software and used in the e-government domain within the 'eGov IT helpdesk support' case. The paper discusses the advantages of the approach and the lessons learned from its first deployment. We strongly conjecture that the approach is noticeable due to its simplicity, low cost, low maintenance effort and its highly customizable architecture. Moreover, it can be used to integrate 'live' and evolving data from multiple sources, without the need to abandon original applications and data stores, if certain guidelines and recommendations, as discussed in the paper, are followed.

Future plans concern the use of S3-AI in other domains where evolving data generated from disparate applications need to be integrated. This will result to further refinements of the approach and further recommendations for its use in different contexts under different requirements. Beyond e-Government, other scenarios of use concern integration of sensor data and data sources for the purposes of marine infor-mation analysis, related to other projects that we are currently involved. The aim is to apply the presented approach in order to open, link and aggregate heterogeneous and distributed 'live' and evolving data (environmental, marine, shipping, geospatial) from multiple sites, offering high-quality services for efficient decision-making in the shipping domain, preserving the original/source applications, datasets and their ownership.

Last but not least, we identify that there is a data security issue to be tackled in the future. For the specific case reported, data have been opened only for inter-organizational use i.e. between distributed sites/departments of NARA. The data we made publicly available in the open Web (via the approach web page) is only dummy/experimental data filtered for demonstration purposes. Secure endpoints are a future concern of ours.

### Acknowledgements

Authors acknowledge the contribution of researchers from the International Hellenic University (IHU), National Technical University of Athens (NTUA), LOV (http://lov.okfn.org), PwC and Univ. of Southampton. Special thanks to G. Santim-patakis, N. Loutas, C. Gutteridge, B. Vatant and P. Vandenbussche.